\documentclass[journal,twocolumn]{IEEEtran}

\usepackage{cite}
\usepackage{amsmath,amssymb,amsfonts}
\usepackage{algorithmic}
\usepackage{graphicx}
\usepackage{longtable}
\usepackage{amsmath}

\begin{document}

\newtheorem{thm}{Theorem}
\newtheorem{lem}{Lemma}
\newtheorem{cor}{Corollary}
\newtheorem{defn}{Definition}

\title{Double-Side Polarization and Beamforming Alignment in Polarization Reconfigurable MISO System with Deep Neural Networks}
\author{\IEEEauthorblockN{Paul Seungcheol Oh, Han Han, Joongheon Kim, Sean Kwon} \\
\thanks{Joongheon Kim and Sean Kwon is the corresponding author.}
}
\maketitle

\vspace{-1cm}
\begin{abstract}
Polarization reconfigurable (PR) antennas enhance spectrum and energy efficiency between next-generation node B (gNB) and user equipment (UE). This is achieved by tuning the polarization vectors for each antenna element based on channel state information (CSI). On the other hand, degree of freedom increased by PR antennas yields a challenge in channel estimation with pilot training overhead. 
This paper pursues the reduction of pilot overhead, and proposes to employ deep neural networks (DNNs) on both transceiver ends to directly optimize the polarization and beamforming vectors based on the received pilots without the explicit channel estimation. 
Numerical experiments show that the proposed method significantly outperforms the conventional first-estimate-then-optimize scheme by maximum of 20\% in beamforming gain.
\end{abstract}

\section{Introduction}
Polarization reconfigurable (PR) antennas are a promising technology that can realize reconfigurable wireless channel with flexible radio frequency (RF) front end to outperform the conventional wireless communication systems \cite{Review_RA_2020, RA_Analysis_2022, Kwon_Molisch_Globecom, Heath_LinearPol, AntennaSelection, Oh_Kwon_MPS_Journal}. 
Studies have been recently conducted to leverage the PR antennas to multi-antenna communication systems. In paticular, 
\cite{Kwon_Molisch_Globecom, Heath_LinearPol, AntennaSelection} utilize PR antennas to remarkably increase the channel capacity in a point-to-point multiple-input-multiple-output (MIMO) system. 
On the other hand, the system design for the optimal polarization vectors depends crucially on the knowledge of the channel state information (CSI). In practice, obtaining the sufficient CSI in a multi-PR-antenna communication system is difficult, since the PR wireless channels typically provide higher degrees-of-freedom than the conventional ones with the corresponding pilot signal design.

This paper considers a double-side polarization and beamforming alignment problem within a PR multiple-input-single-output (PR-MISO) system. In this setting, a single-antenna user equipment (UE) first sends pilots to a multi-antenna next generation node B (gNB), and the gNB designs its polarization and the beamforming vectors based on the received pilots. Subsequently, the gNB sends pilots towards the UE, enabling the UE to design its polarization vectors accordingly. By transmitting and receiving pilots from both transceiver sides, the gNB and UE learn to implicitly coordinate with each other without additional feedback cost. 

The double-side polarization and beamforming alignment problem is very challenging to solve since it involves the estimation of the PR channel which is inherently higher-dimensional than what is directly observed through pilots at both the gNB and UE.
This is because the PR systems exploit the PR antennas based on the channel depolarization matrix \cite{Kwon_Stuber_GeoTheory}; which comes with the cost of increase in the dimensionality of the channel compared to what can be observed through pilots. With such increase in the channel dimension, the channel estimation becomes highly non-trivial at both tranceiver sides; thus, it leads to excessive pilot training overhead as the scale of the multi-antenna PR system grows.

The main idea of this paper is that significant savings on pilots overhead is possible by leveraging deep neural networks (DNNs) to bypass explicit channel estimation at both transceiver sides. 
Specifically, we employ two separate DNNs at the gNB and UE to directly optimize polarization and beamforming vectors at the gNB as well as the optimal polarization vector at the UE directly based on their received pilots. Simulation results show that the proposed end-to-end data-driven approach significantly reduces the pilot overhead compared to the conventional first-estimate-then-optimize scheme. 

\section{PR-MISO Communication System}
\subsection{System Model}
This paper considers a polarization reconfigurable multiple-input-single-output (PR-MISO) communication system, in which a gNB with $N_t$ PR antenna elements configured as a uniform linear array (ULA) supports a UE with a single PR antenna element. Each antenna element at the gNB is equipped with a radio frequency (RF) chain.
\begin{figure}[ht]
  \centering
  \includegraphics[width=0.34\textwidth]{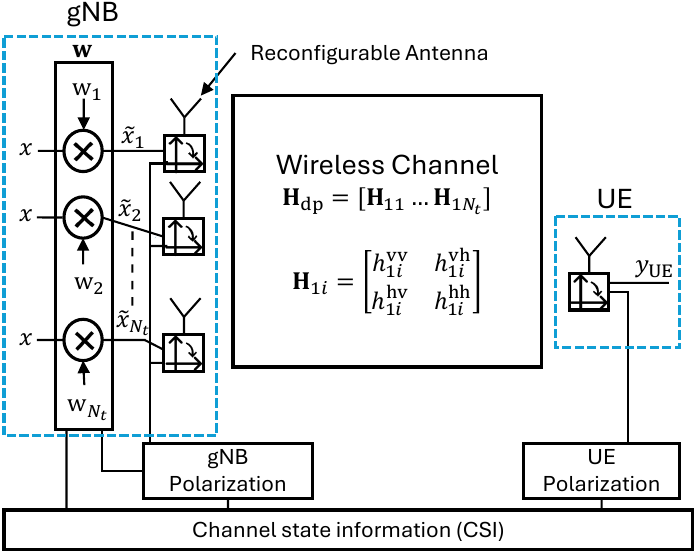} 
  \caption{PR-MISO system.}
  \label{fig:PR-MIMO_system}
\end{figure}
As illustrated in Fig. \ref{fig:PR-MIMO_system}, we denote the \emph{depolarized} channel from the gNB to the UE as $\bold{H}_{\rm dp}$, that captures the channel depolarization effect \cite{Kwon_Stuber_GeoTheory}:
\begin{equation}
    \label{eq:Hup}
    \bold{H}_{\rm dp}=
    \begin{bmatrix}
    \bold{H}_{11} & \hdots & \bold{H}_{1N_t}
    \end{bmatrix} \in \mathbb{C}^{2\times 2N_t}.
\end{equation}
Here, elements $\bold{H}_{1i}$'s represent the \emph{polarization-basis} matrix:
\begin{equation}
\label{eq:PolBasisMatrix}
    \bold{H}_{1i}=
    \begin{bmatrix}
    h_{1i}^{\rm vv} & h_{1i}^{\rm vh} \\
    h_{1i}^{\rm hv} & h_{1i}^{\rm hh}
    \end{bmatrix},
\end{equation}
where $h_{1i}^{\rm \phi\psi}$, $\phi\in\{\rm v,h\}$ and $\psi\in\{\rm v,h\}$ denotes the $\phi\psi$-channel impulse response from the $i^{\rm th}$ $\psi$-polarization gNB antenna element to the $\phi$-polarization UE antenna element.

The primary advantage of PR antenna systems comes from tuning the polarization vectors which gives another degree-of-freedom in the system design. Specifically, we use $\bold{p}_{{\rm gNB},i}$ and $\bold{p}_{\rm UE}$ to represent the gNB polarization vector at $i^{\rm th}$ PR antenna element at the gNB and UE polarization vector, respectively. They are defined as
\begin{align}   
    \label{eq:pol_vec_gNB}
    \bold{p}_{{\rm gNB},i} &=
    \begin{bmatrix}
        \cos{\theta_i}  \\
        \sin{\theta_i}
    \end{bmatrix}
    , ~ i=1,2,\hdots ,N_t,\\ 
    \bold{p}_{\rm UE} &=
    \begin{bmatrix}
        \cos{\theta}_{\rm UE} \\
        \sin{\theta}_{\rm  UE}
    \end{bmatrix},  \label{eq:pol_vec_UE}
\end{align}
where $\theta_{\theta_i,\theta_{\rm UE}}\in[0,\frac{\pi}{2}]$. 
Given (\ref{eq:Hup}), (\ref{eq:pol_vec_gNB}) and (\ref{eq:pol_vec_UE}), we define the effective polarized channel as \cite{Kwon_Molisch_Globecom}:
\begin{equation}
    \label{eq:Heff}
    \bold{H}_{\rm eff}=
    \begin{bmatrix}
        \bold{p}^{\top}_{\rm UE}\bold{H}_{11}\bold{p}_{{\rm gNB},1} & \hdots & \bold{p}^{\top}_{\rm UE}\bold{H}_{1N_t}\bold{p}_{{\rm gNB},N_t}
    \end{bmatrix},
\end{equation}
where each entry consists of polarization vectors of the UE and gNB modifying the \emph{polarization-basis} matrices. 

Let $x\in \mathbb{C}$ denote the data symbol subjected to power constraint $\mathbb{E}[|x|^2]\leq\rho$. Then the received signal at the UE can be expressed as
\begin{equation}
    \label{eq:Received_sig_UE}
    y_{\rm UE}=\underbrace{\bold{p}^{\top}_{\rm UE}\bold{H}_{\rm dp}\bold{P}_{\rm gNB}}_{\bold{H}_{\rm eff}}\bold{w}x+n_{\rm UE}\in \mathbb{C},
\end{equation}
where $\bold{w}\in \mathbb{C}^{N_t}$ denotes the normalized gNB downlink beamforming vector towards UE, i.e., $||\bold{w}||^2=1$. Here, $n_{\rm UE} \sim \mathcal{C}\mathcal{N}(0,\sigma^2_{\rm UE})$ is the additive Gaussian noise at the UE side and $\bold{P}_{\rm gNB}$ is a block diagonal matrix that contains the gNB polarization vectors along its diagonal as:
\begin{equation}
\label{eq:blkdiag}
    \bold{P}_{\rm gNB}={\rm blkdiag}(\bold{p}_{{\rm gNB},1},\hdots,\bold{p}_{{\rm gNB},N_t}) \in \mathbb{R}^{2N_t\times N_t}.
\end{equation}
To this end, we can express the achievable rate $R$ at the UE as \cite{cover1999elements, FundamentalsWC}
\begin{equation}
    \label{eq:channel_capacity}
    R = \log\Bigg(1+\frac{\big|\bold{p}^\top_{\rm UE}\bold{H}_{\rm dp}\bold{P}_{\rm gNB}\bold{w}\big|^2}{\sigma_{\rm UE}^2}\Bigg). 
\end{equation}

The objective of this paper is to jointly design $\bold{w},\bold{p}_{\rm UE}$ and $\bold{P}_{\rm gNB}$ so as to maximize the achievable rate $R$. We note that the design on $\bold{p}_{\rm UE}$ and $\bold{P}_{\rm gNB}$ is equivalent to designing the polarization angles $\theta_{\rm UE}, \theta_1, ..., \theta_{N}$. The optimization problem is formulated as 
\begin{equation}
\label{eq:max_formulation}
\begin{aligned}
\max_{\theta_{\rm UE},\bold{\theta}_1,\hdots,\mathbf{\theta}_{N_t},\bold{w}} \quad & \big|\bold{p}^\top_{\rm UE}\bold{H}_{\rm dp}\bold{P}_{\rm gNB}\bold{w}\big|^2\\
\textrm{s.t.} \quad & ||\bold{w}||^2 = 1,\\
  & (\ref{eq:pol_vec_gNB}), (\ref{eq:pol_vec_UE}).\\
\end{aligned}
\end{equation} 
Problem (\ref{eq:max_formulation}) can be solved with the knowledge of perfect CSI (PCSI) via phase matched beamformer \cite{cover1999elements} and iterative polarization optimization algorithm described in \cite{Kwon_Molisch_Globecom}. However, acquiring CSI in a large-scale PR-MISO system is extremely challenging because both the gNB and UE can only observe pilots that are lower dimensional than $\bold{H}_{\rm dp}$. To solve this issue, we propose a double-side pilot protocol assuming that the system is operating in the time division duplex (TDD) mode with the uplink-downlink channel reciprocity. 

\subsection{Double-side Pilot Protocol}
This section describes a double-side pilot protocol in which the gNB first sends a sequence of $L$ downlink pilots $\{x_{{\rm gNB},l}\}^{L}_{l=1}$, under a power constraint $\mathbb{E}[|x_{{\rm gNB},l}|^2]\leq\rho_{\rm gNB}$, to the UE during a designated pilot frame. Subsequently, in the next pilot frame, the UE sends a sequence of uplink $\{x_{{\rm UE},l}\}^{L}_{l=1}$ pilots, under a power constraint $\mathbb{E}[|x_{{\rm UE},l}|^2]\leq\rho_{\rm UE}$, to the gNB. Without loss of generality, we set $x_{{\rm gNB},l}=\sqrt{\rho_{\rm gNB}}$ and $x_{{\rm UE},l}=\sqrt{\rho_{\rm UE}}$. Owe to channel reciprocity, we denote the channel from the gNB to UE and the channel from UE to gNB as $\bold{H}_{\rm dp}$ and $\bold{H}_{\rm dp}^{H}$, respectively. Then, the $l^{\rm th}$ received pilot at the gNB is described as follows 
\begin{align}
    \label{eq:downlink_pilot}
    \bold{y}_{{\rm gNB},l} &= \bold{P}_{{\rm gNB},l}^\top\bold{H}_{\rm dp}^{H}\bold{p}_{\rm UE}x_{{\rm UE},l}+\mathbf{n}_{{\rm gNB},l} \in \mathbb{C}^{N_t},
\end{align}
where $\bold{p}_{\rm UE}$ is randomly generated and fixed across different $l$ and $\bold{P}_{{\rm gNB},l}$ is randomly generated for each $l$. Here, $\mathbf{n}_{{\rm gNB},l}\sim\mathcal{C}\mathcal{N}(0,\sigma^2_{\rm gNB}\bf{I})$ is the additive white Gaussian noise (AWGN) at the gNB. In an analogous manner, $l^{\rm th}$ received pilots at the UE is described as 
\begin{align}
    y_{{\rm UE},l} &= \bold{p}_{\rm UE}^\top\bold{H}_{\rm dp}\bold{P}_{\rm gNB}\bold{w}_lx_{{\rm gNB},l}+n_{{\rm UE},l} \in \mathbb{C}, \label{eq:uplink_pilot}
\end{align}
where $\bold{p}_{\rm UE}$ and $\bold{P}_{\rm gNB}$ are randomly generated and fixed for all $l$ and $\bold{w}_l$ is generated randomly for each $l$ and $n_{{\rm UE},l} \sim \mathcal{C}\mathcal{N}(0,\sigma^2_{\rm UE})$ is the AWGN during the pilot stage at the UE side.


\subsection{Problem Formulation}
This paper proposes to design the beamformer $\bold{w}$ and polarization vectors at the gNB (\ref{eq:blkdiag}) and polarization vector at UE (\ref{eq:pol_vec_UE}) based on the received pilots $\bold{Y}_{\rm gNB}=[\bold{y}_{{\rm gNB},1}, \bold{y}_{{\rm gNB},2}, \hdots ,\bold{y}_{{\rm gNB},L}] \in \mathbb{C}^{N_t\times L}$ and $\bold{Y}_{\rm UE}=[y_{{\rm UE},1}, y_{{\rm UE},2}, \hdots ,y_{{\rm UE},L}]\in \mathbb{C}^{L}$.
The problem of interest is then described as 
\begin{equation}
\label{eq:max_reformulation}
\begin{aligned}
\max_{\mathcal{F}(.), ~ \mathcal{G}(.)} \quad & \mathbb{E}\bigg[\big|\bold{p}^{\top}_{\rm UE}\bold{H}_{\rm dp}\bold{P}_{\rm gNB}\bold{w}\big|^2\bigg]\\
\textrm{s.t.} \quad & ||\bold{w}||^2 = 1, \\ 
\quad &   \{\bold{w},\theta_1,\hdots,\theta_{N_t}\}=\mathcal{F}(\bold{Y}_{\rm gNB}),\\ 
\quad & \theta_{\rm UE}=\mathcal{G}(\bold{Y}_{\rm UE}),
\end{aligned}
\end{equation}
where $\mathcal{F}: \mathbb{C}^{N_t\times L} \rightarrow \mathbb{C}^{N_t} \times \mathbb{R}^{N_t}$ and $\mathcal{G}: \mathbb{C}^{L} \rightarrow \mathbb{R}$. Note that (\ref{eq:max_reformulation}) is a challenging variational optimization problem in which the objective is maximized over the high dimensional functional mapping $\mathcal{F}$ and the functional mapping $\mathcal{G}$. As a solution, we propose to learn $\mathcal{F}$ and $\mathcal{G}$ directly through DNNs based on received pilots.


\section{Least Square Solution for Channel Estimation}
This section describes the conventional first-estimate-then-optimize approach to solve (\ref{eq:max_formulation}) with double-side pilot protocol via least square (LS) channel estimation. We first concatenating downlink pilots into a matrix to form $\bold{Y}_{\rm UE} =[y_{{\rm UE},1}, y_{{\rm UE},2}, \hdots ,y_{{\rm UE},L}]^\top \in \mathbb{C}^{L\times 1}$, expressed as
\begin{equation}
\label{eq:stackedDownlink}
    \bold{Y}_{\rm UE}=\bold{W}^\top\bold{P}^\top_{\rm gNB}\bold{H}_{\rm dp}^\top\bold{p}_{\rm UE} + \bold{n}_{\rm UE},
\end{equation}
with $\bold{W}=[\bold{w}_1,\bold{w}_2, \hdots ,\bold{w}_L] \in \mathbb{C}^{N_t \times L}$, being the matrix of concatenated downlink beamforming vectors used in each pilot frame and 
$\bold{n}_{\rm UE}=[n_{{\rm UE},1},n_{{\rm UE},2},\hdots,n_{{\rm UE},L}]^\top\in \mathbb{C}^{L\times1}$ being the downlink noise vector. With (\ref{eq:stackedDownlink}), we can obtain the estimate of the downlink channel by solving the LS problem formulated as
\begin{equation}
\label{eq:downlink_LS_formulation}
\begin{aligned}
\min_{\bold{H}_{\rm dp}} \quad & \big|\big|\bold{Y}_{\rm UE}-\bold{W}^\top\bold{P}^\top_{\rm gNB}\bold{H}_{\rm dp}^\top\bold{p}_{\rm UE}\big|\big|^2.
\end{aligned}
\end{equation}
The optimal solution to (\ref{eq:downlink_LS_formulation}) is given as 
\begin{equation}
    \label{eq:downlink_LS_solution}
    \hat{\bold{H}}_{\rm dp,{\rm dl}}=\big[(\bold{A}^H\bold{A})^{-1}\bold{A}^H\bold{Y}_{\rm UE}\bold{p}_{\rm UE}^\top(\bold{p}_{\rm UE}\bold{p}_{\rm UE}^\top)^{-1}\big]^\top,
\end{equation}
where $\bold{A}\triangleq\bold{W}^T\bold{P}^T_{\rm gNB}$ is effective gNB receiving sensor. 

In a similar manner, we formulate the uplink pilots as 
\begin{equation}
\begin{aligned}
    \bar{\bold{Y}}_{\rm gNB}&=[\bold{y}_{{\rm gNB},1}^\top, \bold{y}_{{\rm gNB},2}^\top, \hdots ,\bold{y}_{{\rm gNB},L}^\top]^\top\in \mathbb{C}^{LN_t\times1} \\
    \bar{\bold{Y}}_{\rm gNB}&= \bar{\bold{P}}_{\rm gNB}^\top\bold{H}_{\rm up}^\top\bold{p}_{\rm UE}+\bold{N}_{\rm gNB},
\end{aligned}
\end{equation}
where $\bar{\bold{P}}_{\rm gNB}=[\bold{P}_{{\rm gNB},1},\bold{P}_{{\rm gNB},2},\hdots,\bold{P}_{{\rm gNB},L}]\in\mathbb{R}^{2N\times LN_t}$ is a concatenated matrix of (\ref{eq:blkdiag}) in each pilot frame and $\bold{N}_{\rm gNB}=[\mathbf{n}_{{\rm gNB},1}, \mathbf{n}_{{\rm gNB},2}, \hdots, \mathbf{n}_{{\rm gNB},L}]$ is the concatenated uplink noise matrix. To this end, uplink LS problem for estimation of $\bold{H}_{\rm up}$ is formulated as, 
\begin{equation}
\label{eq:uplink_LS_formulation}
\begin{aligned}
\min_{\bold{H}_{\rm dp}} \quad & \big|\big|\bar{\bold{Y}}_{\rm gNB}-\bar{\bold{P}}_{\rm gNB}\bold{H}_{\rm dp}^\top\bold{p}_{\rm UE}\big|\big|^2.
\end{aligned}
\end{equation}
The optimal solution to (\ref{eq:uplink_LS_formulation}) is 
\begin{equation}
\label{eq:uplink_LS_solution}
    \hat{\bold{H}}_{\rm dp,{\rm ul}}=\big[(\bar{\bold{P}}_{\rm gNB}\bar{\bold{P}}_{\rm gNB}^\top)^{-1}\bar{\bold{P}}_{\rm gNB}\bar{\bold{Y}}_{\rm gNB}\bold{p}_{\rm UE}^\top(\bold{p}_{\rm UE}\bold{p}_{\rm UE}^\top)\big]^\top.
\end{equation}

By utilizing the estimated channels on both ends, the gNB can design the optimal polarization and beamforming vector using (\ref{eq:uplink_LS_solution}), while UE can also design its optimal polarization vector with (\ref{eq:downlink_LS_solution}). Given that there are $4N_t$ unknown parameters in $\bold{H}_{\rm dp}$ and the gNB receives $N_t$ dimensions with each pilot signal, four uplink pilots are sufficient to recover the channel if the noise is not present. In contrast, the UE observes only a single dimension per pilot signal and requires $4N_t$ pilots to recover the channel. This leads to increase in pilot overhead. 

\section{Proposed Deep Learning Solution}
\begin{figure}[t]
  \centering
  \includegraphics[width=0.43\textwidth]{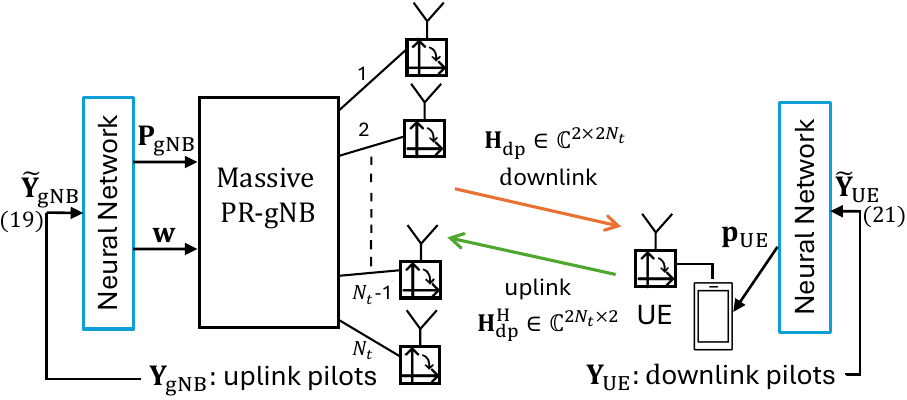} 
  \caption{Proposed DNN based PR-MISO system}
  \label{fig:PR-MIMO_system with DNN}
\end{figure}

The goal of the proposed DNN framework is to maximize the beamforming gain in (\ref{eq:max_reformulation}) from $\bold{Y}_{\rm gNB}$ and $\bold{Y}_{\rm UE}$ directly based on the received pilots so as to bypass explicit channel estimation. As shown in Fig. \ref{fig:PR-MIMO_system with DNN}, the mapping functions $\mathcal{F}$ and $\mathcal{G}$ are learned separately at both sides with their respective DNNs. 
Specifically, $\mathcal{F}$ is parameterized by a DNN with the size $2LN_t\times h_1 \times \hdots \times h_j \times 3N_t$, where it has $j$ hidden layers with each consisting of $h_j$ neurons. Particularly, 
the input of the network is described as
\begin{equation}
    \tilde{\bold{Y}}_{\rm gNB} = \big[\text{vec}(\Re\{\bold{Y}_{\rm gNB}\})^\top, \text{vec}(\Im\{\bold{Y}_{\rm gNB}\})^\top\big] \in \mathbb{C}^{2LN_t},
\end{equation}
which is a concatenated and flattened vector consisting of real and imaginary parts of $\bold{Y}_{\rm gNB}$. 

The output of the network is separated and passed to sigmoid and normalization layers to produce $\bold{P}_{\rm gNB}$ and $\bold{w}$. Let $\bold{z}\in\mathbb{R}^{3N_t}$ denote the output vector of the gNB's DNN and define $[\bold{z}]_{m_1:m_2}$ as subvector of $\bold{z}$ index from $m_1$ to $m_2$. To generate the polarization angles, we first extract $[\bold{z}]_{1:N_t}$ and process it through a sigmoid layer to produce $\theta_i= \sigma([\bold{z}]_{1:N_t})\cdot \frac{\pi}{2}$, where $\sigma(x) = \frac{1}{1 + e^{-x}}$ is the sigmoid function. This yields $N_t$ polarization angles, which are used to construct $\bold{P}_{\rm gNB}$ according to (\ref{eq:pol_vec_gNB}). Next, we handle the remaining $2N_t$ output of gNB's DNN, i.e., $[\bold{z}]_{N_t+1:3N_t}$. These components are processed through a normalization layer defined by
\begin{equation}
    \label{eq:normalization layer}
    \bold{w}=\frac{[\bold{z}]_{N_t+1:2N_t}+j[\bold{z}]_{2N_t+1:3N_t}}{\big|\big|[\bold{z}]_{N_t+1:3N_t}\big|\big|_{2}}.
\end{equation} This normalization layer satisfies the power constraint of the downlink beamformer and produces $\bold{w}$.

In a similar way, UE's mapping $\mathcal{G}$, is parameterized by UE's DNN with the size  $2L\times h_1 \times \hdots \times h_j \times 1$. The input of UE's DNN is described as
\begin{equation}
    \tilde{\bold{Y}}_{\rm UE}=\big[{\rm vec}(\Re\{\bold{Y}_{\rm UE}\})^\top, {\rm vec}(\Im\{\bold{Y}_{\rm UE}\})^\top\big]\in \mathbb{C}^{2L}. 
\end{equation}
Let $\bold{g}\in\mathbb{R}$ denote output of UE's DNN. We process this through the sigmoid layer, described as $\theta_{\rm UE}=\sigma(\bold{g})\cdot\frac{\pi}{2}$, to produce the UE polarization angle. This gets employed to construct $\bold{p}_{\rm UE}$ with (\ref{eq:pol_vec_UE}). With the sigmoid and normalization layers at gNB's side and a sigmoid layer at UE's side, the network is trained in unsupervised manner with loss function set as $-\mathbb{E}\big[\big|\bold{p}^\top_{\rm UE}\bold{H}_{\rm dp}\bold{P}_{\rm gNB}\bold{w}\big|^2\big]$.

\section{Simulation Results}
\label{sec:results}
\subsection{Simulation Setup}
The evaluation of the proposed DNN framework takes into account the massive PR-MISO system model. The gNB equipped with $N_t=64$ PR antenna elements supports the UE with a single PR antenna element. We simulate a scenario where there are only non-line of sight (NLoS) paths between the gNB and UE with rich scatterings. Therefore, each entry of polarization-basis matrix is modeled as 
$h^{\rm \phi\psi}_{1i}\sim\mathcal{CN}(0,1)$, where $\rm \phi\in\{v,h\}$, $\rm \psi\in\{v,h\}$. This is a more challenging scenario compared to the e.g., Rician fading channels, since no obvious channel structures such as line-of-sight (LoS) component can be learned by the neural network. 
We employ a 3-layer $512\times512\times3N_t$ DNN and $512\times512\times1$ DNN for the gNB and UE, respectively. 
The DNNs are trained and tested with PyTorch \cite{PyTorch}. 
\begin{figure}[t]
  \centering
  \includegraphics[width=0.424\textwidth,height=0.30\textwidth]
  {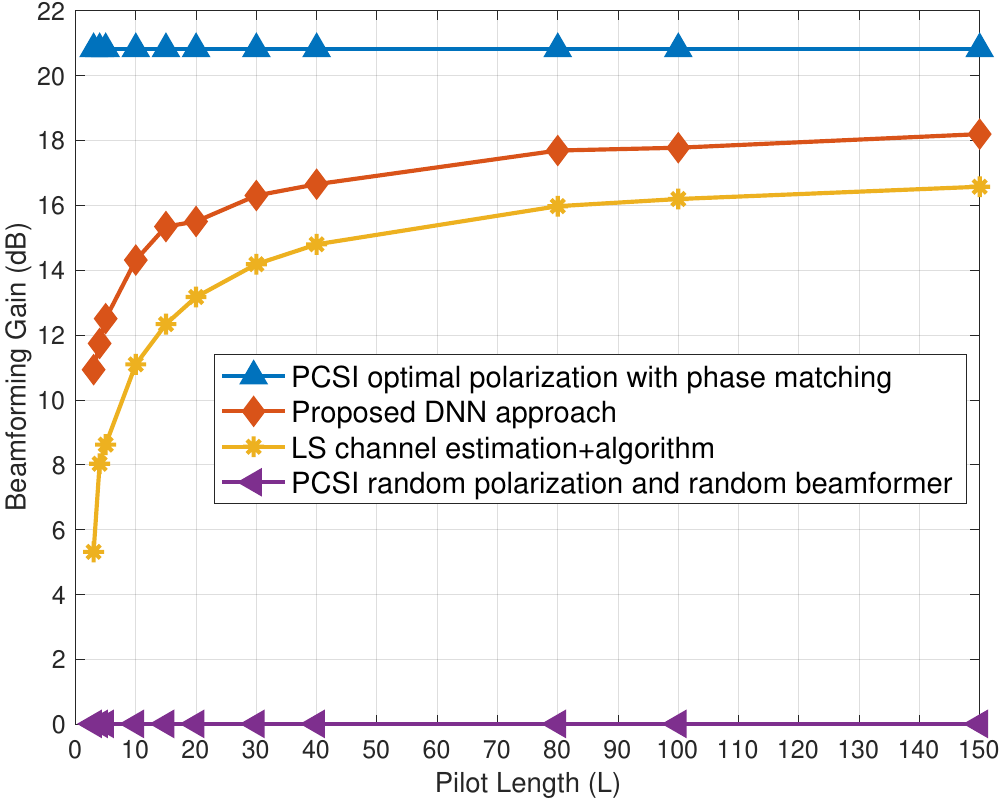} 
  \caption{Beamforming gain vs. pilot length with SNR=-10 dB}
  \label{fig:bfVsPilot}
\end{figure}

We compare the result of the trained DNNs in testing stage with three benchmarks. First, PCSI with optimal polarization vectors found with iterative polarization optimization algorithm \cite{Kwon_Molisch_Globecom} and phased matched beamforming vector. Regardless of pilot length this gives the highest beamforming gain, because it is given the PCSI. Second, optimal polarization vectors and phased matched beamforming vector with LS estimated channel. Lastly, random polarization and beamforming vectors. This outputs the worst beamforming gain because every parameter is set to random. Each point over the curve is averaged over 10000 realization of the channel. 

\subsection{Results}
\begin{figure}[t]
  \centering
  \includegraphics[width=0.42\textwidth,height=0.30\textwidth]{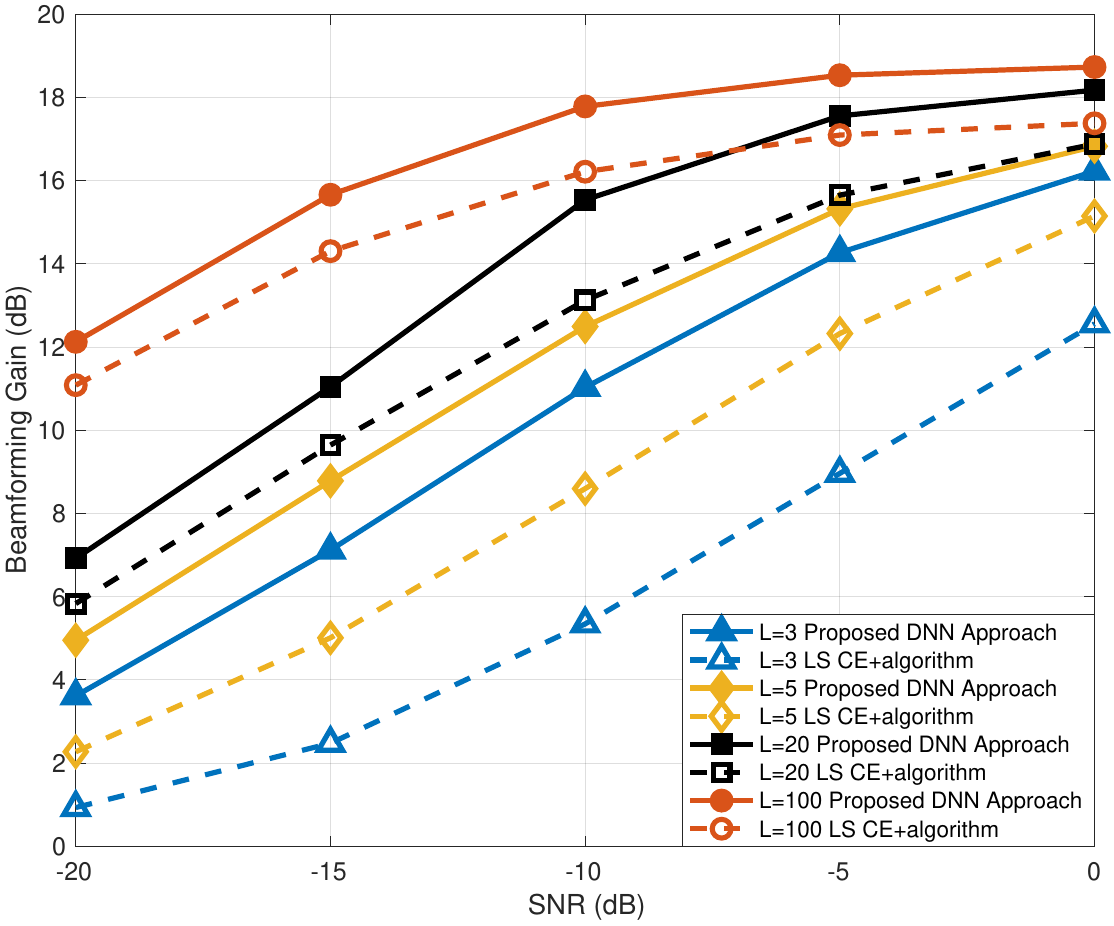} 
  \caption{Beamforming gain vs. SNR}
  \label{fig:BFvsSNR}
\end{figure}

We emphasize the performance gain from the proposed DNN approach to that of the conventional first-estimate-then-optimize scheme as the pilot length increases in Fig. \ref{fig:bfVsPilot}. First, remark that the DNN approach always results higher beamforming gain than that of the gain computed with LS channel estimation. Note that when pilot length is 3, there is $20\%$ gain from the proposed method over the LS solution. However, the gap decreases to about $8\%$ gain from LS to proposed result as the pilot length goes over 4. This is because fully digital gNB require 4 pilots to sufficiently estimate $4N_t$ unknown parameters of the channel. This point is clearly emphasized in Fig. \ref{fig:BFvsSNR}. As illustrated, in all SNR regimes, the performance gap between the proposed approach and LS approach is the highest out of all values of $L$ when $L=3$. This gives us an intuition that the proposed method will perform even superior when number of RF chain is limited at the gNB. Observation of such scenario will be left for our future work. Further, the proposed approach when $L=5$ achieves similar performance as LS solution when $L=20$. Similarly, when $L=20$, it outperforms LS solution when $L=100$ at SNR$=-5,0$ dB. 

\section{Conclusion}
The PR-MISO system takes advantage of the additional degrees-of-freedom of polarization by effectively exploiting the channel depolarization matrix. However, conventional first-estimate-then-optimize scheme confronts a challenge in channel estimation, due to the required higher dimensional channel estimation from the lower dimensional pilot measurements. This paper resolves the challenge via proposing the employment of DNNs at both transceiver sides to learn optimal polarization and beamforming vectors at the gNB and polarization vectors at the UE based on received pilots. Simulation results illustrate that the proposed method of bypassing explicit channel estimation significantly outperforms the conventional first-estimate-then-optimize scheme.
\bibliographystyle{IEEEtran}
\bibliography{references}

\end{document}